\makeatletter
\@ifundefined{@parse@version@dash}{%
\def\@parse@version#1{\@parse@version@0#1}
\def\@parse@version@#1/#2/#3#4#5\@nil{%
\@parse@version@dash#1-#2-#3#4\@nil}
\def\@parse@version@dash#1-#2-#3#4#5\@nil{%
  \if\relax#2\relax\else#1\fi#2#3#4 }
}{}
\makeatother

\documentclass[aps,prx,twocolumn]{revtex4-2}

\usepackage{graphicx}
\usepackage{commath}
\usepackage{color}
\usepackage[percent]{overpic}
\usepackage{titlesec}
\usepackage{multirow}
\usepackage{hyperref}
\usepackage{bbold}
\usepackage{pdfcomment}
\usepackage[table]{xcolor}

\usepackage{cleveref}

\usepackage{bm}

\definecolor{MyBlue}{RGB}{0,0,220}
\definecolor{MyRed}{RGB}{214,39,40}

\begin{document}

\title{Predicting any small-scale statistics of high-Reynolds-number turbulence using ensemble simulations}

\author{Lukas Bentkamp}
\email{lukas.bentkamp@uni-bayreuth.de}
    \affiliation{Theoretical Physics I, University of Bayreuth, Universit\"atsstr.\ 30, 95447 Bayreuth, Germany}
\author{Michael Wilczek}
    \email{michael.wilczek@uni-bayreuth.de}
    \affiliation{Theoretical Physics I, University of Bayreuth, Universit\"atsstr.\ 30, 95447 Bayreuth, Germany}

\date{\today}

\begin{abstract}
The complex small-scale statistics of turbulence are a result of the combined cascading dynamics through all scales of the flow. Predicting these statistics using fully resolved simulations at the high Reynolds numbers that typically occur in engineering, geophysical, and astrophysical flows will exceed the capabilities of even the largest supercomputers for the foreseeable future. A common observation is that high-Reynolds-number flows are organized in clusters of intense turbulent activity separated by large regions of quiescent flow. We here show that small-scale statistics in high-Reynolds-number turbulence can be predicted based on an ensemble hypothesis, stating that they can be emulated by the statistical mixture of a heterogeneous ensemble of lower-Reynolds-number simulations. These simulations are forced at smaller scales with energy injection rates varying across the ensemble. We show that our method predicts complex gradient statistics from the recent literature, including the joint QR-PDF and extreme dissipation and enstrophy statistics, to unprecedented accuracy. Remarkably, we find that the weight distribution needed for the ensemble method to make predictions can be inferred from the anomalous scaling exponents of turbulence. Thus combining theory with fully resolved simulations, our method can be readily applied to predict a wide range of statistics at high Reynolds number but low cost, while opening up various avenues for further theoretical and numerical exploration.

\end{abstract}
 
\maketitle
\section{Introduction}
\begin{figure*}
\includegraphics{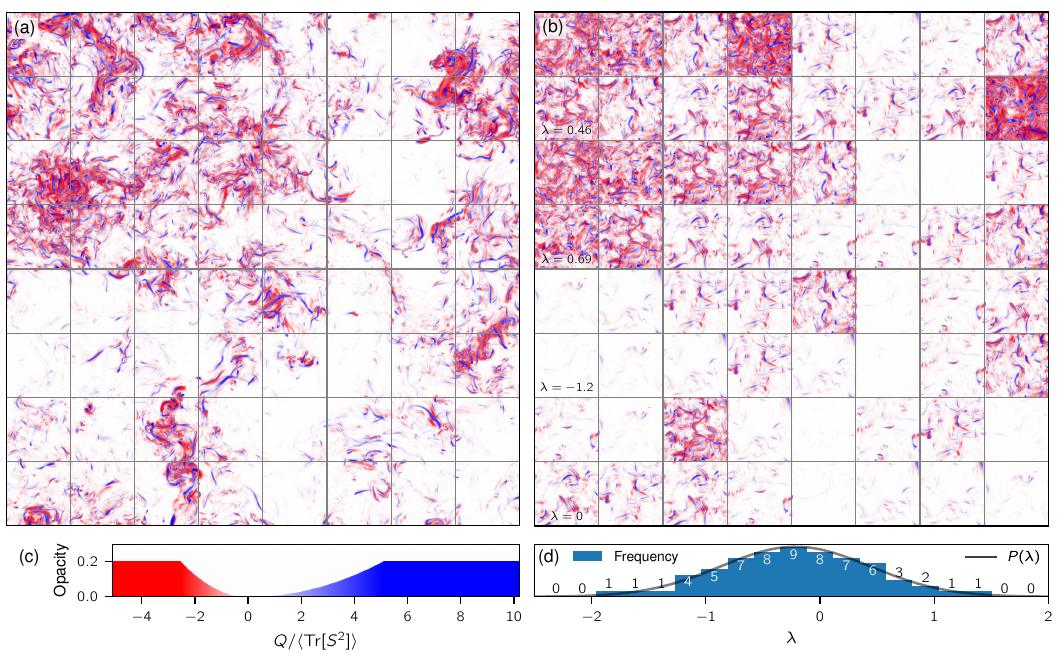}%
\caption{Illustration of the ensemble hypothesis. Panels (a) and (b) are volume renderings~\cite{Schroeder2006} of extreme values of the gradient invariant $Q = -\frac12 \mathrm{Tr}[A^2]$ from DNS, with high enstrophy events in blue and high dissipation events in red. (a) Slab of our reference simulation at $R_\lambda= 284$ on a $2048^3$ grid. (b) Arrangement of slabs from the ensemble simulations \texttt{ens256} (see table~\ref{tab:ensemble_sims}) on $256^3$ grids. Each ensemble member corresponds to a different energy injection rate $e^\lambda \overline{\varepsilon}$ encoded by the logarithmic parameter~$\lambda$ (some of the values labeled in the plot). One central ensemble member corresponding to $\lambda=0$ at $R_{\lambda} = 63$ has the same energy injection rate $\overline{\varepsilon}$ as the reference simulation. (c) Color code used for the volume rendering, with the normalization taken only from the reference simulation and used for all panels. (d) Histogram of the frequency of each tile (bar chart and digits) as displayed in panel (b), approximating the probability given by the Gaussian distribution for $\lambda$ from our theory (black line).  The visual similarity between panels (a) and (b) suggests that the ensemble simulations may reproduce the statistics of the larger simulation. Note that this visualization shows quite a small range of $\lambda$ values compared to the full ensemble (which ranges between $-5.1 \leq \lambda \leq 5.8$) due to the fact that we can draw only 64 samples. The spatial arrangement of the tiles in panel (b) was chosen to match the snapshot in panel (a) visually. \label{fig:mosaic}}
\end{figure*}

Most turbulent flows relevant to applications can reach Reynolds numbers far beyond the range that can be simulated in direct numerical simulations (DNS) of the Navier-Stokes equation. For example, atmospheric turbulence can have Taylor-scale Reynolds numbers $R_\lambda \sim 10^4$~\cite{RisiusAMT2015}. Astrophysical molecular clouds can even reach $R_\lambda \sim 10^6$ or $10^7$~\cite{MieschA1999}. For simulations of turbulence in a periodic box, which is one of the most suitable geometries for modern supercomputers, the highest Reynolds number currently reached is $R_\lambda \sim 2500$~\cite{YeungCPC2025}. For this reason, alternative computational approaches are needed.

For resolving the large-scale flow features, large-eddy simulations are widely used and arguably provide the closest approximation to realistic turbulent flows while still being computationally feasible~\cite{Sagaut2005}. Predicting small-scale flow features, however, still remains the subject of fundamental research. Much more than the large scales, the small scales are a distillate of the entire turbulent flow dynamics, their characteristics resulting from the cascade processes across the scales~\cite{Vela-MartinJFM2022}. Their statistics are skewed~\cite{ChevillardPDNP2006,IshiharaJFM2007,BosPoF2012}, heavy-tailed, and depend highly on Reynolds number~\cite{IshiharaJFM2007,AntoniaJFM2015}. Knowing about the statistics of these small scales is crucial for understanding various phenomena, such as rain formation in turbulent clouds~\cite{ShawARFM2003}, plankton patchiness in the ocean~\cite{AbrahamN1998,DurhamNC2013,BreierPNAS2018}, and the impact of turbulence on combustion~\cite{SreenivasanFTaC2004,HamlingtonPoF2012} and mixing~\cite{GollubPRL1991,Chate2012}.

Some of the complexities of small-scale turbulence statistics can be explained by the spatial flow structure. At the smallest level, turbulence essentially consists of intertwined vorticity filaments~\cite{JimenezJFM1993}. Viewing these at larger and larger scales reveals that they organize in large clusters~\cite{IshiharaJFM2007,IshiharaFTC2013,ElsingaJFM2023,MukherjeePRL2024}, separating the flow into regions of high and low activity (see figure~\ref{fig:mosaic}a). Arguably, the main effect of increasing the Reynolds number can be captured by understanding how the additional large-scale flow structures add variations to the flow. Here, we propose a novel, simple, and extendable methodology to predict a wealth of small-scale statistical quantities at high Reynolds number, by assuming that they can be described as a mixture of statistics from a heterogeneous ensemble of flows at lower Reynolds numbers (see figure~\ref{fig:mosaic}b). 

As of now, obtaining accurate small-scale turbulence statistics requires extracting data directly either from real-world measurements, or from fully resolved DNS. When those are not available, one typically has to resort to phenomenological modeling. A rather successful line of phenomenological models for the small scales are Lagrangian velocity gradient models~\cite{MeneveauARFM2011,JohnsonARFM2024}. These models focus on the dynamics of the velocity gradient tensor of single Lagrangian particles and proved quite successful in capturing its complex statistics, such as the teardrop shape of the joint distribution of the $Q$ and $R$ invariants~\cite{JohnsonJFM2016,JohnsonPRF2017,PereiraJFM2018,LuoPRF2022,BuariaPNAS2023,CarbonePRL2024}. Recent velocity gradient models started focusing also on the intermittency properties of the small scales and capturing Reynolds number dependence~\cite{JohnsonPRF2017,PereiraJFM2018,LuoPRF2022,BuariaPNAS2023}.

A conceptually very different approach is taken by the multifractal framework~\cite{MandelbrotJFM1974,benzi1984,FrischTAP1985,PaladinPR1987,meneveau1987,BoffettaJPAMT2008,BenziJSP2009,Frisch1995}, which emerged from Kolmogorov's and Oboukhov's refined similarity hypotheses~\cite{oboukhov1962,kolmogorov1962}. From the start, various mostly equivalent formulations of the multifractal approach have coexisted, such as the geometric interpretation~\cite{FrischTAP1985,meneveau1987} and probabilistic formulations~\cite{MandelbrotJFM1974,MandelbrotTaNSE1976,MeneveauJFM1991a,MandelbrotPRSLSMPS1997}. 
At its core, the multifractal formalism describes scaling in the inertial range, but it was soon extended to infer viscous-range statistics as well~\cite{PaladinPRA1987,NelkinPRA1990,BenziPRL1991,YakhotPASMaiA2004,YakhotPhysicaD2006}. In fact, some of the more recent Lagrangian velocity gradient models also incorporate concepts from multifractality~\cite{PereiraJFM2018,LuoPRF2022}. While multifractal approaches can produce very good fits to experimental and simulated data~\cite{chevillard2012,PereiraJFM2018}, applying them in practice can be challenging. This is partly because the main postulate of multifractal theory are scaling relations, which need to be supplemented with additional assumptions to yield concrete predictions such as probability density functions (PDFs). One approach to predict PDFs, for example, is to assume some base distributions, which are then mixed probabilistically in a way consistent with multifractal phenomenology~\cite{BenziPRL1991,CastaingPDNP1990,chevillard2012,Sosa-CorreaPRF2019}. 
The method presented here turns out to be similar to this mixing approach, but with a powerful modification: we replace the base distributions by ensembles of small DNS of the Navier-Stokes equation~\cite{BentkampJFM2025}. This eliminates the need for specific assumptions and produces pure multifractal Reynolds number extrapolations of arbitrary statistical quantities.

Our main conceptual contribution is the `ensemble hypothesis'---stating that high-Reynolds-number small-scale statistics can be described as a superposition of statistics of flows at smaller Reynolds numbers---which we combine with the idea of anomalous scaling from the multifractal literature. It turns out that prescribing only a set of anomalous scaling exponents determines entirely how to construct the ensemble superposition. In order to make concrete predictions, we conduct an ensemble of DNS at low Reynolds numbers, with varying energy injection rates across the ensemble. In this sense, our method could be considered a realization of `small-eddy simulations'~\cite{JimenezJoT2003,MoitroPoF2024}. The computational cost of our ensemble simulations remains orders of magnitude lower than the effort needed for turbulence simulations at the high Reynolds numbers we aim for. Since our method is simulation-based, it can predict (at basically any Reynolds number) any small-scale statistics that can be measured in a simulation, including, e.g., the complex statistics of the velocity gradient tensor. It does so to excellent accuracy, rivalled only by recent machine learning approaches~\cite{BuariaPNAS2023,CarbonePRL2024}. Thanks to its simplicity, our approach may also inspire various applications and extensions of the ensemble simulation paradigm.

\section{The ensemble hypothesis}
The basis of our method is the ensemble hypothesis. With increasing Reynolds number, turbulence phenomenology tells us that every active scale of the flow adds spatio-temporal fluctuations to the cascade, which accumulate toward the smaller scales~\cite{Frisch1995}. This is consistent with the observation that high-Reynolds-number flows develop large-scale clusters of turbulent activity~\cite{IshiharaJFM2007,IshiharaFTC2013,ElsingaJFM2023}. Based on this idea, we hypothesize that we can effectively capture small-scale statistics of a high-Reynolds-number flow by mixing the statistics from a heterogeneous ensemble of low-Reynolds-number flows with varying energy injection rates.

We here consider simulations of statistically stationary, homogeneous, and isotropic turbulence with fixed energy injection rate at the large scales. Given a fixed forcing scheme and geometry, the statistics of these flows only depend on three basic physical properties: the energy injection rate $\overline{\varepsilon}$, the integral scale $L$ at which the energy is injected, and the kinematic viscosity $\nu$. Together, they can be used to compute the Reynolds number
\begin{align} \label{eq:Re_definition}
  \mathrm{Re} = C_\varepsilon^{-1/3} \nu^{-1} (\overline{\varepsilon} L^4)^{1/3} \,,
\end{align}
where $C_\varepsilon$ is a non-dimensional quantity converging to a constant at large enough Reynolds number~\cite{SreenivasanTPoF1984,SreenivasanPoF1998,McCombPRE2015}.
We write the corresponding ensemble average for the statistics of each of these flows as $\langle \,\cdot\, ;\, L, \overline{\varepsilon}, \nu\rangle$. For the present work, we keep the viscosity constant across all simulations (both high and low Reynolds numbers). In our simulations, we can choose the injection rate $\overline{\varepsilon}$ explicitly and the integral scale $L$ is fixed by choosing a forcing wavenumber band.

Using the introduced notation, we can now formally state the ensemble hypothesis for a generic small-scale quantity $X$ as
\begin{align} \label{eq:ensemble_hypothesis}
  \left\langle X ;\, e^Z L, \overline{\varepsilon}, \nu \right\rangle &= \int \dif\lambda\, P(\lambda; Z) \left\langle X ; \, L, e^\lambda \overline{\varepsilon}, \nu \right\rangle\,.
\end{align}
Here, the left-hand side of the equation corresponds to the reference simulation that we want to emulate, and the right-hand side corresponds to a superposition of the statistics of the ensemble simulations. There is a factor $e^Z$ between the integral scale of the reference simulation and the one of the ensemble simulations, where we call $Z \geq 0$ the scale gap.
The ensemble hypothesis~\eqref{eq:ensemble_hypothesis} is the assertion that the additional large-scale structures due to the scale gap cause small-scale intermittency that can be modeled by mixing ensemble simulations with varying energy injection rates. These variations are described by the parameter $\lambda$, whose distribution $P(\lambda; Z)$ still needs to be determined. For $Z = 0$, the equality can be satisfied trivially by $P(\lambda; 0) = \delta(\lambda)$.

We choose to write the injection rate coefficient as $e^\lambda$ to account for the multiplicative nature of the energy cascade. With this choice, the well-known log-normal model for the turbulent cascade~\cite{oboukhov1962,kolmogorov1962} corresponds to a Gaussian distribution for $\lambda$ (as we show in the next sections). Energy redistribution factors like $e^\lambda$ appear in various forms in the multifractal literature, such as Mandelbrot's weighted curdling model~\cite{MandelbrotTaNSE1976,Frisch1995}, but typically without asserting far-reaching statistical equivalence as we do in~\eqref{eq:ensemble_hypothesis}.

As an example for the application of~\eqref{eq:ensemble_hypothesis}, we consider the dissipation rate $X=\varepsilon$. Since our flows are statistically stationary, the mean dissipation rate equals the injection rate, $\langle \varepsilon; L, \overline{\varepsilon}, \nu \rangle = \overline{\varepsilon}$. Inserted into~\eqref{eq:ensemble_hypothesis}, this implies
\begin{align}
  \overline{\varepsilon} &= \int \dif\lambda\, P(\lambda; Z) e^\lambda \overline{\varepsilon} \\
  \Rightarrow 1 &= \int \dif\lambda\, P(\lambda; Z) e^\lambda\,, \label{eq:mean_injection_constraint}
\end{align}
which constrains the distribution of $\lambda$.

It is important to note, however, that the ensemble hypothesis allows to make much more general predictions. For example, we can predict entire probability density functions (PDFs) by using their definitions as averages over delta functions. For example, the single-point longitudinal gradient PDF for a flow with integral scale $L$, injection rate $\overline{\varepsilon}$, and viscosity $\nu$ can be defined as
\begin{align}
  \mathrm{PDF}(A_{11}; L, \overline{\varepsilon}, \nu) &= \left\langle \delta\left(\partial_1 u_1 - A_{11}\right); L, \overline{\varepsilon}, \nu \right\rangle\,,
\end{align}
where $\partial_1 u_1$ is the realization of the gradient and $A_{11}$ is the corresponding sample space variable. The ensemble hypothesis~\eqref{eq:ensemble_hypothesis} then implies that
\begin{align} \label{eq:pdf_superposition}
  \mathrm{PDF}(A_{11}; e^Z L, \overline{\varepsilon}, \nu) &= \int \dif\lambda\, P(\lambda; Z) \mathrm{PDF}(A_{11}; L, e^{\lambda}\overline{\varepsilon}, \nu)\,.
\end{align}
This means that PDFs of larger-Reynolds-number flows can be written as superpositions of PDFs of smaller-Reynolds-number flows.

In principle, the distribution $P$ of the injection rate variations $\lambda$ could depend on various flow properties including the Reynolds number. However, in the following section, we show that the distribution of $\lambda$ can be determined entirely by assuming that the model has to display anomalous scaling. We find that it is a function of the scale gap $Z$ only. Moreover, we find an explicit formula for the $\lambda$ distribution given the presumably universal anomalous scaling exponents of turbulence.

\section{Relation to anomalous scaling} \label{sec:relation_to_scaling_exponents}
It has been established that various statistical quantities in turbulence display anomalous scaling~\cite{PaladinPR1987,Frisch1995}, i.e., they follow power laws with non-trivial scaling exponents. Among others, this is the case for moments of longitudinal velocity increments $\delta_\ell u$ at scale $\ell$ (these moments are called structure functions), for moments of longitudinal velocity gradients $\partial_1 u_1$, and for moments of the dissipation rate $\varepsilon$. The corresponding scaling exponents $\zeta_p$, $\rho_m$, and $d_n$ are defined by the scaling relations~\cite{Frisch1995,SchumacherNJP2007}
\begin{subequations}
\begin{align}
  \left\langle (\delta_\ell u)^p\right\rangle &\sim \ell^{\zeta_p} \\
  \left\langle (\partial_1 u_1)^m\right\rangle &\sim \mathrm{Re}^{\rho_m} \propto \nu^{-\rho_m} \label{eq:def_rhom} \\
  \left\langle \varepsilon^n \right\rangle &\sim \mathrm{Re}^{d_n} \propto \nu^{-d_n}\,.
\end{align}
\end{subequations}
The Reynolds number scaling is meant at constant large scales, i.e., the relations can be written as an inverse proportionality to the viscosity $\nu$. The remaining dependence on the large-scale quantities $L$ and $\overline{\varepsilon}$ can be inferred by matching dimensions,
\begin{subequations}
\begin{align}
  \left\langle (\delta_\ell u)^p ; L, \overline{\varepsilon}, \nu  \right\rangle &= A_{p} (\overline{\varepsilon} L)^{p/3} \left(\ell / L\right)^{\zeta_p} \label{eq:scaling_equation_1}\\
  \left\langle (\partial_1 u_1)^m ; L, \overline{\varepsilon}, \nu  \right\rangle &= B_{m} (\overline{\varepsilon} L^{-2})^{m/3} \nu^{-\rho_m} (\overline{\varepsilon} L^4)^{\rho_m/3} \label{eq:scaling_equation_2}\\
  \left\langle \varepsilon^n ; L, \overline{\varepsilon}, \nu  \right\rangle &= C_{n}\, \overline{\varepsilon}^n  \nu^{-d_n} (\overline{\varepsilon} L^4)^{d_n/3} \,. \label{eq:scaling_equation_3}
\end{align}
\end{subequations}
We write these as equations to emphasize that the remaining non-dimensional proportionality constants $A_{p}$, $B_{m}$, and $C_{n}$ are assumed universal. They only depend on the type of forcing. However, it is important to note that the relations hold only asymptotically at high Reynolds number. Additionally, for the increments, $\ell$ needs to lie in the inertial range.

We want to combine \cref{eq:scaling_equation_1,eq:scaling_equation_2,eq:scaling_equation_3} with the ensemble hypothesis~\eqref{eq:ensemble_hypothesis}. Hence we need to assume that the Reynolds numbers of all flows involved (both reference and ensemble simulations) are large enough such that scaling holds and that they have an inertial range in which $\ell$ lies. Then we can use~\eqref{eq:ensemble_hypothesis} with $X = (\delta_\ell u)^p,\, (\partial_1 u_1)^m,\, \varepsilon^n$, yielding
\begin{widetext}
\begin{subequations}
\begin{align}
  (\overline{\varepsilon} e^Z L)^{p/3} \left(\ell e^{-Z} / L\right)^{\zeta_p}
  &= \int\dif\lambda\, P(\lambda; Z)\left(e^\lambda \overline{\varepsilon} L\right)^{p/3} \left(\ell / L\right)^{\zeta_p} \label{eq:wide_ens_hypothesis_1}\\
  \left(\overline{\varepsilon} e^{-2Z} L^{-2}\right)^{m/3} \nu^{-\rho_m} \left(\overline{\varepsilon} e^{4Z} L^4 \right)^{\rho_m/3}
   &= \int\dif\lambda\, P(\lambda; Z)\left(e^\lambda \overline{\varepsilon} L^{-2}\right)^{m/3} \nu^{-\rho_m} \left(e^\lambda \overline{\varepsilon} L^4\right)^{\rho_m/3} \label{eq:wide_ens_hypothesis_2} \\
   \overline{\varepsilon}^n \nu^{-d_n} (\overline{\varepsilon} e^{4Z} L^4)^{d_n/3}
   &= \int\dif\lambda\, P(\lambda; Z)(e^{\lambda}\overline{\varepsilon})^n  \nu^{-d_n} (e^\lambda \overline{\varepsilon} L^4)^{d_n/3} \,. \label{eq:wide_ens_hypothesis_3}
\end{align}
\end{subequations}
\end{widetext}
Remarkably, all instances of $\overline{\varepsilon}$, $\nu$, $\ell$, and $L$ in these equations cancel. Upon taking the logarithm, this leaves only
\begin{subequations}
\begin{align}
  \left(\frac{p}{3} - \zeta_p \right) Z
  &= K\left( \frac{p}{3}; Z \right) \label{eq:exponents_KZ_1} \\
  \left(-\frac{2m}{3} + \frac{4\rho_m}{3}\right) Z &= K\left( \frac{m}{3} + \frac{\rho_m}{3} ; Z\right) \label{eq:exponents_KZ_2}\\
  \frac{4d_n}{3} Z &= K\left(n + \frac{d_n}{3}; Z\right) \,, \label{eq:exponents_KZ_3}
\end{align}
\end{subequations}
where
\begin{align} \label{eq:cumulant_generating_function}
  K(t; Z) = \log \int\dif\lambda\,P(\lambda; Z) e^{\lambda t}
\end{align}
is the cumulant generating function of the $\lambda$ distribution. Each of the equations~\labelcref{eq:exponents_KZ_1,eq:exponents_KZ_2,eq:exponents_KZ_3} determines $K$ as a function of the scaling exponents and of $Z$. Presupposing that the scaling exponents are universal constants, we keep $Z$ as the only parametric dependence. Since a distribution is determined by its cumulant generating function (if it exists), we can also write $P(\lambda; Z)$ with the only parametric dependence being $Z$, as we stated earlier. In particular, the distribution $P(\lambda; Z)$ does not depend explicitly on the Reynolds number of any of the flows.

Many connections can be drawn to similar expressions in the multifractal literature (e.g., refs.~\cite{MandelbrotTaNSE1976,MuzyPRE2002,YakhotPhysicaD2006}). As is typical of multifractal approaches, too, predictions of our method are based on the knowledge of a set of scaling exponents. In our case, the equations \labelcref{eq:exponents_KZ_1,eq:exponents_KZ_2,eq:exponents_KZ_3} are formulas (one explicit and two implicit) for the computation of the cumulant generating function $K$ of the $\lambda$ distribution, given a set of scaling exponents and a scale gap $Z$. As the scale gap is simply related to the targeted Reynolds number, the scaling exponents remain as the only non-trivial input to the model besides the ensemble simulations. For example, we can compute $K(t; Z)$ through~\eqref{eq:exponents_KZ_1} given a parameterization of the $\zeta_p$ exponents that can be found in the literature~\cite{meneveau1987,ShePRL1994,YakhotPRE2001,sreenivasan2021}. Cumulant generating functions are convex, which is consistent with the known concavity of the $\zeta_p$ exponents~\cite{Frisch1995}. The corresponding PDF $P(\lambda; Z)$ can be computed from $K(t; Z)$ by an inverse two-sided Laplace transform applied to $\exp(K(t; Z))$ (analogous, e.g., to the inverse Mellin transform in ref.~\cite{YakhotPhysicaD2006}). Then, the ensemble hypothesis~\eqref{eq:ensemble_hypothesis} specifies how to compute concrete high-Reynolds-number statistics.

\begin{figure}
  \includegraphics{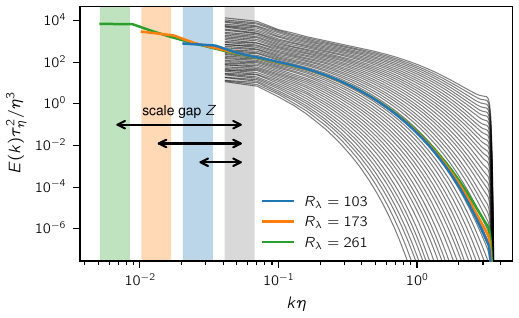}
  \caption{Energy spectra of the reference simulations (colored, solid lines) compared to the main ensemble simulations \texttt{ens256} (grey, solid lines). The ensemble simulations are all forced at the same scale indicated by the grey shaded area. The reference simulations are forced at larger scales, with their forcing bands indicated by the colored shaded areas. The ratio of forcing scales is quantified by the scale gap $Z$. Across the ensemble, the energy injection rate is varied. Since we are normalizing by the Kolmogorov scales $\eta$ and $\tau_\eta$ of the reference simulations, the energy spectra of the different ensemble members shift vertically. \label{fig:energy_spectra}}
\end{figure}

Given the analogies, our ensemble hypothesis could be considered an interpretation of the multifractal conjecture~\cite{BenziJSP2009}. In fact, by design, all of our method's Reynolds number extrapolation is consistent with the anomalous scaling exponents. A specific feature of our $\lambda$ distribution, however, is that it models only the intermittency of the `scale gap' between the ensemble and the reference flow (visualized in figure~\ref{fig:energy_spectra}). All the smaller-scale dynamics and statistics is fully resolved in our ensemble DNS. As a result, our method is ready to make predictions of concrete small-scale statistics such as PDFs without manually specifying anything more than the scaling exponents $\zeta_p$. 

Before moving on to setting up the ensemble, let us remark on some theoretical implications of the ensemble hypothesis combined with anomalous scaling. Looking at the relations \labelcref{eq:exponents_KZ_1,eq:exponents_KZ_2,eq:exponents_KZ_3}, we find that any cumulant generating function $K(t; Z)$ consistent with anomalous scaling has to be linear in the scale gap $Z$. Translating this into the $\lambda$ distribution, we find that, by a convolution theorem, doubling $Z$ corresponds to a convolution of the $\lambda$ distribution with itself, i.e.,
\begin{align} \label{eq:convolution_property}
  P(\lambda; 2Z) = P(\lambda; Z) \ast P(\lambda; Z)\,.
\end{align}
In fact, in order to exist for all $Z \geq 0$, $P(\lambda; Z)$ needs to be an infinitely divisible distribution~\cite{NovikovPRE1994,MuzyPRE2002,ChainaisEPJB2006}.

Moreover, somewhat surprisingly, the relations \labelcref{eq:exponents_KZ_1,eq:exponents_KZ_2,eq:exponents_KZ_3} allow to derive relations between the different sets of scaling exponents. If a single cumulant generating function $K(t; Z)$ is supposed to consistently satisfy all three relations for all values of $p$, $m$, and $n$, then the exponents need to relate to each other in a certain way. In particular, given a fixed value of $m$ we consider $p=m+\rho_m$. Then, we can equate \eqref{eq:exponents_KZ_1} with \eqref{eq:exponents_KZ_2} yielding
\begin{align} \label{eq:nelkin_rhom}
  \rho_m = m - \zeta_{m + \rho_m} \,.
\end{align}
This is the well-known Nelkin relation, originally derived by extrapolating the multifractal description of the inertial range to fluctuating dissipation scales~\cite{NelkinPRA1990,Frisch1995}. This original derivation requires a full multifractal description in terms of singularity spectra. Here, we require the more observational assumption of scaling~\labelcref{eq:scaling_equation_1,eq:scaling_equation_2,eq:scaling_equation_3}, which, combined with the ensemble hypothesis~\eqref{eq:ensemble_hypothesis}, leads to the Nelkin relation~\eqref{eq:nelkin_rhom}.

The same can be done for the dissipation exponents. We choose $p=3n+d_n$ and equate~\eqref{eq:exponents_KZ_1} with \eqref{eq:exponents_KZ_3}, which yields
\begin{align} \label{eq:nelkin_dn}
  d_n &= n - \zeta_{3n+d_n}\,,
\end{align}
consistent with previous literature, too~\cite{BorgasPTRSLSPES1997,BoschungPRE2015}. The fact that these results from the multifractal literature come out as a direct consequence of the ensemble hypothesis~\eqref{eq:ensemble_hypothesis} reaffirms that the ensemble idea is consistent with the multifractal approach. Moreover, the ensemble hypothesis seems to provide a simple mechanism to derive multifractal relations, which is a promising theoretical path to explore.

\section{Description of the ensemble simulations}
With the above considerations, we are able to construct a predictive model for small-scale statistics at high Reynolds number using ensemble simulations. At the heart of this model is a set of small direct numerical simulations of the Navier-Stokes equation. The simulated flows need to be approximately statistically homogeneous, isotropic, and stationary. We need to be able to set the energy injection rate in these simulations, so we choose a forcing that amplifies the velocity Fourier coefficients in a large-scale wavenumber band at every time step, such that the energy injection rate is fixed to a prescribed value. In order to comply with the right-hand side of~\eqref{eq:ensemble_hypothesis}, all simulations across the ensemble have the same forcing wavenumber band and the same viscosity. The energy injection rate varies across the ensemble. One `central' simulation in the ensemble (corresponding to $\lambda=0$) has the same energy injection $\overline{\varepsilon}$ as the flows that we want to describe (the reference simulations). The other ensemble simulations have injection rates $e^\lambda \overline{\varepsilon}$ for a range of linearly spaced $\lambda$ values. The range of $\lambda$ values needed to make predictions depends on the target Reynolds number and on the statistics of interest. 

\begin{table}
  \caption{Sets of ensemble simulations. $n$ is the number of member simulations of each ensemble. $R_{\lambda,\mathrm{ens}}$ is the Taylor-scale Reynolds number of the central ensemble member ($\lambda=0$). As we keep the viscosity constant across all simulations, it is the different physical box sizes (and proportionally to that, the forcing scales) that lead to different central Reynolds numbers. Here, we mainly present predictions based on the \texttt{ens256} ensemble. For figure~\ref{fig:extreme_gradients}, we replace its member simulations with the largest $\lambda$ values by $512^3$ simulations, doubling the effective small-scale resolution (in brackets). \label{tab:ensemble_sims}}
  \begin{ruledtabular}
  \begin{tabular}{lllllll}
  name & grid size & $n$ & $R_{\lambda,\mathrm{ens}}$ & $k_M\eta$ & box size & $\Delta k$ \\ \hline\\[-1em]
  \texttt{ens128} & $128^3$ & 48 & 36 & 3.0 & $(2\pi)^3$ & $1$\\
  \texttt{ens256} & $256^3$ ($512^3$) & 48 & 63 & 3.0 (6.0) & $(4\pi)^3$ & $1/2$\\
  \texttt{ens512} & $512^3$ & 48 & 103 & 3.0 & $(8\pi)^3$  & $1/4$
  \end{tabular}
  \end{ruledtabular}
\end{table}
\begin{table}
  \caption{Reference simulations. In order to evaluate our method, we compare its predictions with the statistics from these three simulations. While their viscosity (and energy injection rate) is the same as in the (central) ensemble simulations, their physical box sizes and forcing scales are larger. $(t_1-t_0)/T$ denotes the simulation length without transient, measured in integral times $T$. Then, we additionally compare to data from the literature~\cite{BuariaNJP2019,BuariaPRL2022,BuariaPNAS2023,BuariaPRF2025} with much higher Reynolds number than these. \label{tab:reference_simulations} }
  \begin{ruledtabular}
  \begin{tabular}{llllll}
  grid size & $R_{\lambda, \mathrm{ref}}$ & $k_M\eta$ & box size & $\Delta k$ & $(t_1-t_0)/T$ \\ \hline\\[-1em]
  $512^3$ & 103 & 3.0 & $(8\pi)^3$ & $1/4$ & $34$\\
  $1024^3$ & 173 & 3.0 & $(16\pi)^3$ & $1/8$ & $18$\\
  $2048^3$ & 261 & 3.0 & $(32\pi)^3$ & $1/16$ & $5.5$
  \end{tabular}
  \end{ruledtabular}
\end{table}
The different sets of ensemble simulations we use in this work are listed in table~\ref{tab:ensemble_sims}. For most of the data presented here, we use \texttt{ens256}, an ensemble of 48 simulations on a $256^3$ grid with injection rates corresponding to $\lambda \in [-5.1, 5.8]$ spaced with $\Delta\lambda = 0.23$. These individual simulations have Reynolds numbers $21 \leq R_\lambda \leq 180$, with the central $(\lambda=0)$ simulation at $R_{\lambda,\mathrm{ens}} = 63$. Without transient, they last between $18$ and $24$ integral time scales. They are forced in a wavenumber band $k \in [1.4\Delta k, 2.3\Delta k]$, where $\Delta k$ is the resolution in wavenumber space. The time step is chosen such that the temporal resolution criterion is $\mathrm{CFL} = 0.2$ across all simulations. The spatial resolution of the central ensemble simulation is $k_M\eta=3.0$, where $k_M$ denotes the maximum resolved wavenumber and $\eta$ the Kolmogorov length scale. Measured with respect to their individual average dissipation rates, the spatial resolution of the other simulations varies between $0.7 \leq k_M\eta \leq 10.6$. However, the high-injection-rate simulations corresponding to low spatial resolution are only used to predict rare, high-gradient events. If one considered only these high-gradient events in the reference simulations, one would see that they face the same issues of limited spatial resolution as in the ensemble simulations. For a fair comparison of the spatial resolution, we need to compute the Kolmogorov scale from the dissipation rate averaged over the entire ensemble, weighted by the distribution $P(\lambda; Z)$. Due to the constraint~\eqref{eq:mean_injection_constraint}, this is equal to the Kolmogorov scale of the central ensemble simulation (the one with $\lambda = 0$), meaning that the effective spatial resolution of our ensemble is $k_M\eta=3.0$.
\begin{figure}[t]
  \includegraphics{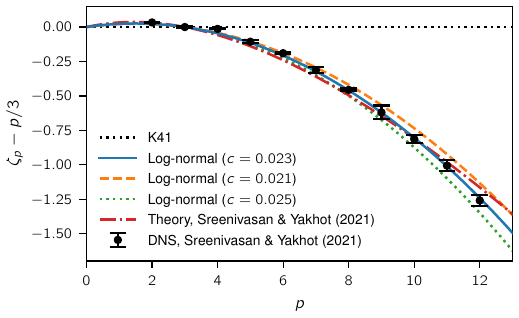}%
  \caption{Anomalous scaling exponents $\zeta_p$ for moments of longitudinal velocity increments. The values for classic Kolmogorov theory~\protect\cite{kolmogorov1941a} (K41), $\zeta_p = p/3$, are subtracted. Recent DNS data from Sreenivasan and Yakhot~\cite{sreenivasan2021} (black dots) is compared to the log-normal parameterization~\eqref{eq:quadratic_scaling_exponents} with different parameters $c$ and the theoretical curve from ref.~\cite{sreenivasan2021}. The log-normal curve for $c=0.023$ (solid, blue) matches the DNS data very well.
  \label{fig:scaling_exponents}}
\end{figure}

In figure~\ref{fig:energy_spectra}, we show the energy spectra of the ensemble simulations (grey lines) compared to the reference simulations (colored lines, described in the next section). All curves are normalized by the same Kolmogorov scales $\eta$ and $\tau_\eta$, computed from the viscosity $\nu$ (that is the same across all simulations) and the reference injection/mean dissipation rate $\overline{\varepsilon}$. All ensemble simulations are forced at the same wavenumber band indicated by the grey, shaded area. Their energy spectra shift vertically due to varying energy injection, but they have identical maximum resolved wavenumbers.

\begin{figure*}[ht]
  \includegraphics{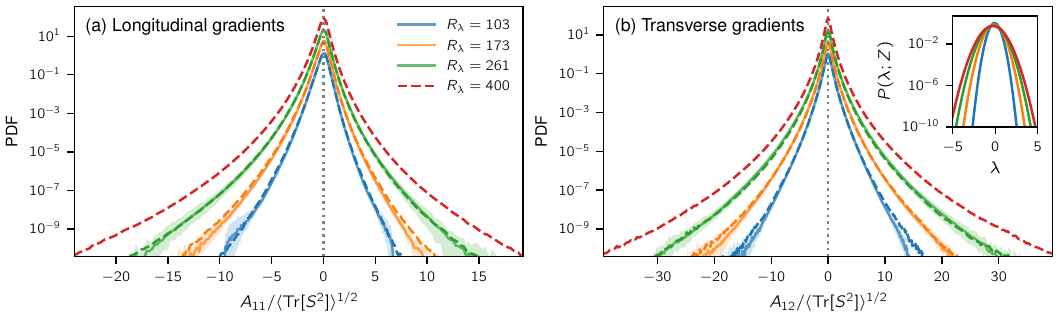}%
  \caption{PDFs of longitudinal (a) and transverse (b) velocity gradients from our reference simulations (solid lines) compared to the model predictions (dashed lines) based on the \texttt{ens256} ensemble and quadratic scaling exponents~\eqref{eq:quadratic_scaling_exponents} with $c=0.023$. They are shifted vertically for clarity. The shaded areas represent confidence intervals for the reference data computed by bias-corrected and accelerated bootstrapping~\cite{VirtanenNM2020}. Bootstrapping was performed over PDF values computed from non-overlapping temporal windows of length equal to one integral time scale. The inset shows the $\lambda$ distributions corresponding to the four target Reynolds numbers. All of them are Gaussian PDFs (parabolic curves) but vary in mean and variance according to~\eqref{eq:gaussian_lambda_pdf}. \label{fig:gradient_pdfs}}
  \end{figure*}
In order to use the ensemble to make predictions, we additionally need a weight distribution $P(\lambda; Z)$. In section~\ref{sec:relation_to_scaling_exponents}, we explained how $P(\lambda; Z)$ can be determined for all $Z$ given a prescribed set of inertial-range scaling exponents $\zeta_p$.  We find that a simple quadratic spectrum of exponents (the log-normal model),
\begin{align} \label{eq:quadratic_scaling_exponents}
  \zeta_p = p/3 - c(p^2-3p)/2\,,
\end{align}
with $c = 0.023$ (close to the value chosen in ref.~\cite{chevillard2012}) produces very good predictions for all of our data. Figure~\ref{fig:scaling_exponents} shows that this parameterization is actually in excellent agreement with recent DNS data for the $\zeta_p$ exponents~\cite{sreenivasan2021}. While it is known theoretically that the $\zeta_p$ exponents for a bounded field have to be monotonous~\cite{Frisch1995} (inconsistent with the quadratic form~\eqref{eq:quadratic_scaling_exponents}), we find that this parameterization suffices for the scope of this work. We note, however, that refined parameterizations of the $\zeta_p$ exponents can be used just as well within our model framework (see section~\ref{sec:modeling_choices}), with the only complication of having to invert the two-sided Laplace transform~\eqref{eq:cumulant_generating_function}.

For the quadratic spectrum of $\zeta_p$ exponents~\eqref{eq:quadratic_scaling_exponents}, the weight distribution can be computed analytically. According to~\eqref{eq:exponents_KZ_1}, we get
\begin{align}
  K\left(t; Z\right) &= \frac{9c(t^2-t)}{2} Z\,.
\end{align}
This is the cumulant generating function of a normal distribution with mean $\mu = -9cZ/2$ and variance $\sigma^2 = 9cZ$, i.e.,
\begin{align} \label{eq:gaussian_lambda_pdf}
  P(\lambda; Z) &= \frac{1}{\sqrt{18\pi cZ}} \exp\left(-\frac{(\lambda + 9cZ/2)^2}{18cZ}\right)\,.
\end{align}
This is all that is needed to make predictions. We associate each ensemble simulation with a $\lambda$ bin, whose probability weight can be computed by integrating the PDF~\eqref{eq:gaussian_lambda_pdf} over this bin. Then any statistical average can be computed as a weighted sum across the ensemble.

\section{Benchmarking the method against DNS data}
In order to assess the performance of our method, we conducted a range of reference simulations, which are listed in table~\ref{tab:reference_simulations}. In order to be able to apply the ensemble hypothesis~\eqref{eq:ensemble_hypothesis} directly (without need for rescaling), all of the reference simulations have the same energy injection rate $\overline{\varepsilon}$ and viscosity $\nu$ (and thus the same Kolmogorov scales). The reference simulations differ in that their physical box size increases with Reynolds number (as listed in table~\ref{tab:reference_simulations}). The forcing scale (and accordingly, the integral length scale) is kept constant relative to the box size, which causes the increase in Reynolds number with the box size. In other words, the forcing wavenumber band is the same multiple of the wavevector discretization $\Delta k$ for all simulations. Specifically, the forcing is applied to all wavenumbers $k \in [1.4\Delta k, 2.3\Delta k]$. These forcing bands are shown as colored shaded areas in figure~\ref{fig:energy_spectra}. The energy spectra of the reference simulations (figure~\ref{fig:energy_spectra}, colored lines) lie on top of each other, with differences only at the large scales. We can also read off the scale gap $Z$ in that figure as the logarithmic distance between the forcing bands of the ensemble and reference simulations. For \texttt{ens256}, it ranges from $Z = \log(2)$ for $R_{\lambda, \mathrm{ref}} = 103$ up to $Z = \log(8)$ for $R_{\lambda, \mathrm{ref}}=261$.

The predictions of gradient component statistics by the ensemble \texttt{ens256} are compared to the reference flows in figure~\ref{fig:gradient_pdfs}. The predictions from our method are simply computed as a superposition of the ensemble simulation PDFs as exemplified in~\eqref{eq:pdf_superposition}. The model reproduces very well the increasingly heavy-tailed PDFs of the reference flows. For the longitudinal gradients, the skewed shape of the distributions is captured as well, a property that typically proves difficult to model~\cite{LavalPoF2001,ChevillardPDNP2006,Sosa-CorreaPRF2019}. The Reynolds number dependence of the model is encoded purely by the dependence of the mean and variance of the Gaussian distribution~\eqref{eq:gaussian_lambda_pdf} on $Z$. These $\lambda$ distributions are shown in the inset of figure~\ref{fig:gradient_pdfs}.
\begin{figure}[ht]
  \includegraphics{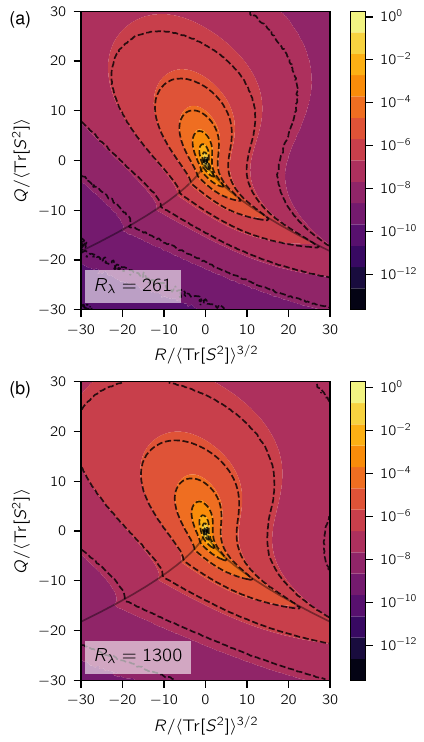}%
  \caption{Joint PDFs of the $Q$ and $R$ invariants of the velocity gradient tensor from reference simulations (colored level sets) compared to our method based on the \texttt{ens256} ensemble and quadratic scaling exponents~\eqref{eq:quadratic_scaling_exponents} with $c=0.023$ (black, dashed iso-contours). Panel (a) uses our largest simulation as reference, panel (b) uses reference data from Buaria \& Sreenivasan~\cite{BuariaPNAS2023}. In order to illustrate the Reynolds number dependence, the same color code and figure ranges are used in both plots.
  \label{fig:qr_pdfs}}
\end{figure}

It is worth noting that the value of $Z$ can also be estimated easily from the Reynolds number alone. To this end, we assume $C_\varepsilon$ constant (compare eq.~\eqref{eq:Re_definition}, refs.~\cite{SreenivasanTPoF1984,SreenivasanPoF1998,McCombPRE2015}) and estimate its value from a fit to all ensemble simulations with $\lambda \geq 0$. Knowing $C_\varepsilon$, we can infer the value of $Z$ for any target Reynolds number $R_{\lambda,\mathrm{ref}}$, and we have
\begin{align} \label{eq:Rlambda_Z_relation}
  R_{\lambda,\mathrm{ref}} \propto e^{2Z/3}\,.
\end{align}
As a result, we can make model predictions for arbitrary Reynolds numbers without ground truth data. To illustrate this, in figure~\ref{fig:gradient_pdfs}, we also show the ensemble prediction for $R_\lambda = 400$, which naturally extrapolates the trend of increasing intermittency and skewness with Reynolds number.

\begin{figure*}[ht]
  \includegraphics{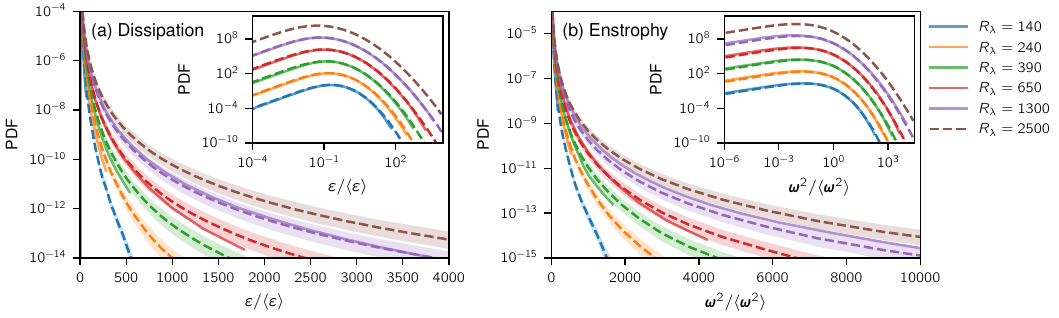}%
  \caption{PDFs of extreme dissipation (a) and enstrophy (b) from refs.~\cite{BuariaNJP2019,BuariaPRL2022} based on their simulations on up to $18\;432^3$ grid nodes (solid) compared to our method (dashed) using the \texttt{ens256} ensemble with the quadratic spectrum~\eqref{eq:quadratic_scaling_exponents} and parameter $c=0.023$. In order to match the high spatial resolution of the reference data, all ensemble simulations with $\lambda \geq 3.2$ are replaced by identical simulations with twice the spatial resolution (grid size $512^3$). The shaded areas correspond to predictions based on values of the parameter $0.021 \leq c \leq 0.025$. The insets show the full range of the PDFs in a log-log representation (vertically shifted for clarity). We also show the model prediction for the additional Reynolds number $R_\lambda = 2500$ to illustrate that no further input is needed for our method to extrapolate to even higher Reynolds number.}
    \label{fig:extreme_gradients}
\end{figure*}
One of the most exciting features of our method is that the ensemble simulations provide entire turbulent fields, from which arbitrary statistics can be computed. A first step into this direction of more complex statistical quantities can be taken by considering the joint distribution of the $Q$ and $R$ invariants of the velocity gradient tensor $A_{ij}$, which are defined as~\cite{JohnsonARFM2024}
\begin{align}
  Q = -\frac12 \mathrm{Tr}[A^2], \quad R = -\frac13 \mathrm{Tr}[A^3]\,.
\end{align}
In figure~\ref{fig:qr_pdfs}a, we compare the QR-PDF of our largest reference simulation at $R_\lambda = 261$ (shown by colored level sets) to the corresponding prediction of our method based on the \texttt{ens256} ensemble (black, dashed lines). The method reproduces the complex teardrop shape of the PDF quantitatively, even in the far tails of the distribution which are rarely reported in the literature.

\section{Extrapolating to the frontier of current turbulence simulations} \label{sec:buaria_data}
It turns out that our ensemble method robustly extrapolates to predicting statistics even at some of the largest Reynolds numbers currently achieved by DNS. For the $QR$-PDF in figure~\ref{fig:qr_pdfs}b, we compare to data from Buaria \& Sreenivasan~\cite{BuariaPNAS2023}, obtained by simulating the Navier-Stokes equation on $12\,288^3$ grid nodes, reaching $R_\lambda= 1300$ at resolution $k_M\eta = 3.0$. As in the other panel, colored level sets show the full DNS data and black, dashed lines are the model predictions. In order to illustrate the differences between the two Reynolds numbers, we use the same color code for the level sets in units of $\langle \mathrm{Tr}[S^2] \rangle$ ($S_{ij}$ is the strain rate tensor) in both panels. The method extrapolates very well up to that Reynolds number to (presumably) unprecedented accuracy, with only minor deviations in the tails. To our knowledge, no other turbulence model allows to predict these small-scale statistics this accurately without fine-tuning of model parameters. In our case, only a set of scaling exponents $\zeta_p$ is needed as input besides the ensemble simulations, and here we achieve excellent agreement with the basic quadratic parameterization~\eqref{eq:quadratic_scaling_exponents} with $c=0.023$.

Using a related set of simulations, the work by Buaria \& Pumir~\cite{BuariaPRL2022} (first data shown in ref.~\cite{BuariaNJP2019}) focused on the far tails of the PDF of enstrophy $\bm\omega^2$ ($\bm\omega$ is the vorticity) and dissipation $\varepsilon = 2\nu \mathrm{Tr}[S^2]$. These tails characterize the probability of the rarest, most extreme events in the dynamics of the velocity gradients. Observing these events in fully resolved simulations at the considered Reynolds numbers requires massive runs occupying entire national-scale supercomputers. In this case, their simulations were performed on up to $18\,432^3$ grid nodes, reaching $R_\lambda= 1300$ at resolution $k_M\eta = 4.5$ and resolution $k_M\eta = 6.0$ for all Reynolds numbers up to $R_\lambda=650$.
Their DNS data is represented by solid lines in figure~\ref{fig:extreme_gradients}. As in their paper, we here show a range of the PDFs up to values $10^3$ to $10^4$ times larger than the mean. Our model predictions using the \texttt{ens256} ensemble are shown as dashed lines. In this case, however, since we are considering extreme gradient events, the spatial resolution becomes a relevant factor. In order to match the high spatial resolution of the reference simulations ($4.5 \leq k_M\eta \leq 6.0$), we replace all those ensemble members with $\lambda \geq 3.2$ (which contribute most to this prediction) by simulations with grid size $512^3$, doubling the effective resolution to be $k_M \eta = 6.0$. As above, we use the simple quadratic spectrum of exponents~\eqref{eq:quadratic_scaling_exponents} with $c=0.023$ and estimate $Z$ from the Reynolds number by extrapolation at constant $C_\varepsilon$.

Indeed, our method can extrapolate even the most extreme ranges of the PDF. Compared to the previous results, we observe some more deviations, as these tails are more sensitive to the limitations of our modeling. The tails can be associated with high-order moments of the gradients. An accurate prediction requires clean power-law scaling of these moments with Reynolds number, where the actual exponents need to match our parameterization. To illustrate the impact of the choice of exponents, the shaded areas show the model predictions for values of the parameter $c$ between $0.021$ and $0.025$, which adds significant variations to the prediction. Moreover, differences between theirs and our simulation schemes may lead to deviations. Nonetheless, again, to our knowledge there is no other turbulence model that allows to predict the tails of these distributions this accurately without fine-tuning of parameters (compare machine learning approach in ref.~\cite{BuariaPNAS2023}).
Also note that the model prediction is not restricted to the tail. As the insets of figure~\ref{fig:extreme_gradients} show, our method produces a prediction of the entire range of the PDFs, with excellent agreement in the bulk. In order to emphasize that the Reynolds number dependence is fully captured by our method, we also show the prediction for $R_\lambda=2500$, which naturally continues the trend of increasing probability of extreme events of the other curves. In the appendix, we show results for a truncated $\lambda$ distribution, which improves the agreement considerably.

Given that our method requires an ensemble of DNS instead of a single large simulation that would be needed for simulating high-Reynolds-number flow, one can ask how much computational advantage there is to the approach. Indeed, in order to make predictions about the extremely rare events that lead to the PDF tails shown in figure~\ref{fig:extreme_gradients}, such events need to be observed at high enough sample number in our ensemble simulations as well. The main computational advantage of our method comes from the fact that the small-scale conditions represented by the different ensemble members occur with vastly different probabilities in the reference simulation. On the one hand, most of the volume of a high-Reynolds-number simulation is occupied by conditions corresponding to energy throughput close to the mean energy injection. In figure~\ref{fig:mosaic}, this is evident in that only a small range of $\lambda$ values ($-2 \leq \lambda \leq 1.5$) is needed to represent the typical conditions of a slice through the reference simulation. These `mean' conditions are strongly oversampled in a high-Reynolds-number simulation, whereas our small ensemble simulations can be stopped when sufficient convergence of statistical quantities is reached. On the other hand, the extremely rare events characterized by the tails in figure~\ref{fig:extreme_gradients} occur quite frequently for our high-$\lambda$ ensemble simulations, but the resulting PDF is then multiplied by a very low value of $P(\lambda; Z)$ to make a prediction. In that sense, the ensemble simulation approach could be considered a form of importance sampling.
\begin{figure}[ht]
  \includegraphics{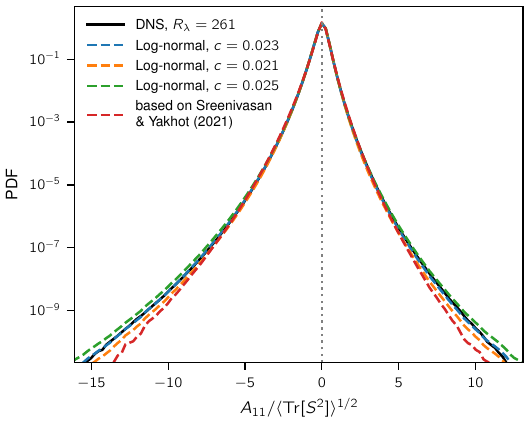}%
  \caption{Model predictions (dashed lines) from \texttt{ens256} ensemble for the longitudinal gradient PDF at $R_\lambda = 261$ using different $\zeta_p$ parameterizations, compared to the reference data (black, solid line). For the prediction based on the $\zeta_p$ parameterization from ref.~\cite{sreenivasan2021} (red, dashed line), we computed the inverse Laplace transform of $\exp(K(t; Z))$ numerically. Depending on the choice of the $\zeta_p$ exponents, the intermittency properties of the predicted PDF change.
  \label{fig:zetap_choices}}
\end{figure}

To make this more concrete, we know~\cite{pope2000} that the required grid size $N^3$ for a certain Reynolds number scales like $N^3 \sim (L/\eta)^3 \sim R_\lambda^{9/2}$. Since we target small-scale statistics, we assume that the reference and ensemble simulations have the same duration in units of the Kolmogorov time. The required number of time steps $M$ for a Kolmogorov time $\tau_\eta$ scales like $M \sim \tau_\eta U/\eta \sim (L/\eta)^{1/3} \sim R_\lambda^{1/2}$ (where $U$ is the root mean square velocity). The overall computational effort scales as $N^3 M \sim (L/\eta)^{10/3} \sim R_\lambda^5$. This means that for the central ensemble simulation we have $N^3 M \sim (L/\eta)^{10/3} \sim R_{\lambda,\mathrm{ens}}^5$ and for each reference simulation we have $N^3 M \sim (e^Z L/\eta)^{10/3} \sim R_{\lambda,\mathrm{ref}}^5$. Overall, we can estimate the computational cost ratio as
\begin{align}
  \frac{\text{Cost full DNS}}{\text{Cost ensemble}} \sim \frac{e^{10Z/3}}{n} \sim \frac{1}{n} \left(\frac{R_{\lambda,\mathrm{ref}}}{R_{\lambda,\mathrm{ens}}}\right)^5\,,
\end{align}
where $n$ is the number of simulations in the ensemble. The required number of simulations $n$ scales approximately with the width of the $\lambda$ distribution, whose variance is linear in $Z$ (see, e.g., \eqref{eq:gaussian_lambda_pdf}), so it is only logarithmic in the Reynolds number (according to~\eqref{eq:Rlambda_Z_relation}). For predicting the statistics from the $R_\lambda = 1300$ simulation in figures~\ref{fig:qr_pdfs} and~\ref{fig:extreme_gradients}, we have $R_{\lambda,\mathrm{ref}}/R_{\lambda,\mathrm{ens}} = 1300/63$ and $n = 48$, implying a factor of computational savings of approximately $7.8\times 10^4$. Here we ignored any differences in spatial resolution and forcing scheme. On top of that, a single set of ensemble simulations with sufficiently wide range of $\lambda$ values can be used to extrapolate to a wide range of different target Reynolds numbers without needing to rerun simulations.

\section{Modeling choices} \label{sec:modeling_choices}
Thanks to its simplicity, our method requires minimal fine-tuning. The main choice that has to be made is the $\zeta_p$ parameterization. In figure~\ref{fig:zetap_choices}, we show how this choice impacts the prediction of the longitudinal gradient PDF at $R_\lambda = 261$. In order to compute $P(\lambda; Z)$ for the $\zeta_p$ parameterization from Sreenivasan \& Yakhot~\cite{sreenivasan2021}, we compute $K(t;Z)$ according to~\eqref{eq:exponents_KZ_1} from the analytical expression of $\zeta_p$. Then, we invert the two-sided Laplace transform~\eqref{eq:cumulant_generating_function} by evaluating the expression at an imaginary argument and computing the Fourier transform numerically (by a fast Fourier transform). The figure shows that the intermittency properties of the prediction are determined by the choice of the scaling exponents. As we go further into the tail of the PDF, higher orders of the $\rho_m$ exponents become relevant, which is responsible for the differences we see between the parameterizations.
\begin{figure*}[ht]
  \includegraphics{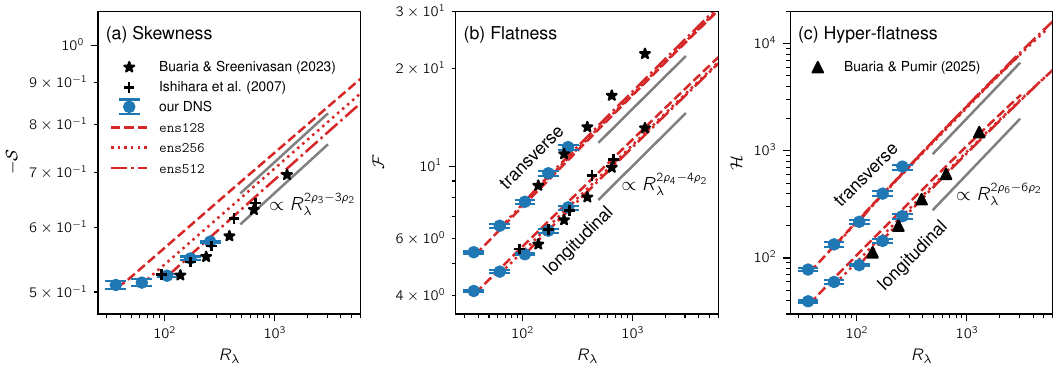}%
  \caption{Normalized moments of gradient components as a function of Reynolds number. (a) Negative skewness of longitudinal gradients. (b) Flatness of both longitudinal and transverse gradients. (c) 6th-order hyper-flatness of both longitudinal and transverse gradients. Our DNS values (blue dots) as well as literature values (Ishihara et al.~\cite{IshiharaJFM2007} as plusses, Buaria \& Sreenivasan~\cite{BuariaPNAS2023} as stars, and Buaria \& Pumir~\cite{BuariaPRF2025} as triangles) are compared to the model predictions using the three ensembles \texttt{ens128} (red, dashed line), \texttt{ens256} (red, dotted line), and \texttt{ens512} (red, dash-dotted line). By design, our method extrapolates by approximate power laws, starting from the value of the central ensemble member. The exponents of the expected power laws (grey, solid lines) can be determined from the $\zeta_p$ parameterization~\eqref{eq:quadratic_scaling_exponents} through the Nelkin relation~\eqref{eq:nelkin_rhom}. Deviations from this scaling are due to the fact that the ensemble simulations themselves do not scale exactly with these exponents, partly due to their low Reynolds number (\texttt{ens128} in panel (a)), partly due to spatial resolution limitations (high Reynolds predictions in panel (c)). Error bars for our DNS values represent confidence intervals from bias-corrected and accelerated bootstrapping~\cite{VirtanenNM2020} over moments computed from non-overlapping temporal windows of length equal to one integral time scale.
  \label{fig:skewness_flatness}}
\end{figure*}

In order to understand what other modeling choices there are that impact the performance of our method, it is illustrative to look at the scaling of skewness, flatness, and 6th-order hyper-flatness of gradient components. They are defined as
\begin{align} \label{eq:def_skewness_flatness}
\begin{gathered}
  \mathcal{S} = \frac{\langle (\partial_1 u_1)^3 \rangle}{\langle (\partial_1 u_1)^2 \rangle^{3/2}}\,,\\
  \mathcal{F} = \frac{\langle (\partial_{1,2} u_1)^4 \rangle}{\langle (\partial_{1,2} u_1)^2 \rangle^2}\,,
  \quad\text{and}\quad \mathcal{H} = \frac{\langle (\partial_{1,2} u_1)^6 \rangle}{\langle (\partial_{1,2} u_1)^2 \rangle^3}
  \,,
\end{gathered}
\end{align}
respectively, where the notation indicates that the gradient components for the (hyper-)flatness can be either longitudinal ($\partial_1 u_1$) or transverse ($\partial_2 u_1$). For the skewness, we consider only longitudinal gradients, because transverse gradient skewness is zero by isotropy. In figure~\ref{fig:skewness_flatness}, we show these quantities computed from our own DNS data in blue, as well as data from Ishihara et al.~\cite{IshiharaJFM2007} (plusses), Buaria \& Sreenivasan~\cite{BuariaPNAS2023} (stars), and Buaria \& Pumir~\cite{BuariaPRF2025} (triangles). The predictions based on the three ensembles listed in table~\ref{tab:ensemble_sims} are shown in red. To compute them, each average in~\eqref{eq:def_skewness_flatness} is written as a superposition according to~\eqref{eq:ensemble_hypothesis}. For example, the skewness prediction is given by
\begin{align}
  \mathcal{S}_\mathrm{pred}(Z) &= \frac{\int\dif\lambda\, P(\lambda; Z) \left\langle (\partial_1 u_1)^3 ; L, e^\lambda \varepsilon_i, \nu\right\rangle}{\left(\int\dif\lambda\, P(\lambda; Z) \left\langle (\partial_1 u_1)^2 ; L, e^\lambda \varepsilon_i, \nu\right\rangle\right)^{3/2}}\,,
\end{align}
where $Z$ is computed from $R_\lambda$ as previously.
Figure~\ref{fig:skewness_flatness} illustrates the various effects that determine the model's performance.

By definition of $\rho_m$~\eqref{eq:def_rhom}, we expect skewness and (hyper-)flatness to scale as
\begin{align}
  \mathcal{S} &\sim \mathrm{Re}^{\rho_3 - 3\rho_2/2} \sim R_\lambda^{2\rho_3 - 3\rho_2}\,, \\
  \mathcal{F} &\sim \mathrm{Re}^{\rho_4 - 2\rho_2} \sim R_\lambda^{2\rho_4 - 4\rho_2}\,,\quad\text{and} \\
  \mathcal{H} &\sim \mathrm{Re}^{\rho_6 - 3\rho_2} \sim R_\lambda^{2\rho_6 - 6\rho_2}\,.
\end{align}
Note that longitudinal and transverse gradients must scale identically within our framework, which is why we do not differentiate between their exponents. If all the assumptions made in section~\ref{sec:relation_to_scaling_exponents} hold, then we expect our model predictions to scale with the above power laws, too. In practice, however, the scaling exponents used to construct the $\lambda$ distribution (which come from the quadratic parameterization~\eqref{eq:quadratic_scaling_exponents}) are neither precisely the same as the ones realized across the ensemble simulations nor the ones in the reference simulations. All discrepancies between these can impact the model prediction.

In order to evaluate this, we compute the theoretical gradient exponents $\rho_m$ from the $\zeta_p$ parameterization~\eqref{eq:quadratic_scaling_exponents} through the Nelkin relation~\eqref{eq:nelkin_rhom}. The expected scaling based on these exponents is shown as grey lines in figure~\ref{fig:skewness_flatness}. The model predictions (red) come close to the slope of the expected scaling, but there are some differences. These are indeed rooted in deviations from the assumptions made for deriving~\labelcref{eq:wide_ens_hypothesis_1,eq:wide_ens_hypothesis_2,eq:wide_ens_hypothesis_3}. We assumed that power-law scaling holds across all simulations, including the ensemble simulations. But figure~\ref{fig:skewness_flatness} shows that scaling becomes apparent only beyond $R_\lambda \sim 100$. This is why specifically the smallest-Reynolds-number ensemble \texttt{ens128} with $R_{\lambda,\mathrm{ens}} = 36$ deviates visibly from the expected scaling (see figure~\ref{fig:skewness_flatness}a) whereas the ensemble \texttt{ens512} with $R_{\lambda,\mathrm{ens}} = 103$ deviates less. A second type of deviation starts to show particularly when predicting the hyper-flatness at high Reynolds numbers: the prediction curves bend slightly downward. We find that this is related to the spatial resolution of the ensembles. Predicting 6th-order hyper-flatness of gradients at $R_\lambda$ beyond $10^3$ requires spatial resolutions higher than the $k_M\eta=3.0$ used for our ensembles. So both of these deviations can be remedied by increasing the size of the ensemble simulations, to higher Reynolds numbers and to higher resolution. While that does increase the computational cost of the ensemble, for a given $Z$ the factor of computational savings between the ensemble and a fully resolved simulation with the same properties remains the same.

Extrapolating with the right scaling exponents as a function of Reynolds number is only one aspect of making an accurate prediction. The other aspect is getting the coefficients of these scaling laws right. The central achievement of our modeling framework is that no assumptions need to be made about the coefficients. Instead, since DNS data is used as input, our method naturally extrapolates starting from the values found in the small-Reynolds-number DNS. This can seen in figure~\ref{fig:skewness_flatness} by the fact that the model scaling laws start from our own DNS data points. In the language of our method, these points correspond to a scale gap $Z=0$, which results in $P(\lambda; Z=0) = \delta(\lambda)$. Hence, the model prediction at that point is simply given by the DNS data from the central ensemble member ($\lambda=0$). Again, this shows that the model performance is significantly improved if the ensemble simulations themselves lie already in the scaling range. For the ensemble \texttt{ens512} starting from $R_{\lambda,\mathrm{ens}} = 103$, we get skewness and flatness predictions that lie closest to the reference data. Nevertheless, also the predictions of the ensemble \texttt{ens256} with $R_{\lambda,\mathrm{ens}} = 63$ are already quite accurate, as we showcased throughout this paper.

Finally, when setting up our method, one needs to choose the range of $\lambda$ values that are covered. We found that the optimal choice depends both on the Reynolds number we want to achieve and on the quantity we are interested in. The further we want to explore the tails of various distributions, the more we need to go to extreme values of $\lambda$. For example, the largest contribution to the prediction of the dissipation PDF shown in figure~\ref{fig:extreme_gradients}a at $\varepsilon/\langle\varepsilon\rangle = 2000$ for $R_\lambda=1300$ comes from the ensemble simulation at $\lambda = 3.93$. The probability mass of the $\lambda$ bin of this ensemble member is $3.1\times 10^{-6}$, so it contributes virtually nothing to the bulk statistics. It is only because most of the other ensemble members do not have any contribution to these extreme values of $\varepsilon$ that this ensemble member becomes relevant. As a general rule, in order to determine the range of $\lambda$ values, one needs to compute the contributions of the ensemble members to the quantity of interest. Based on this, it can be made sure that the ensemble is truncated at a $\lambda$ value with negligible contribution.

\section{Conclusion}
In this work, we put forward the idea that small-scale turbulence at high Reynolds number can be approximated by statistical mixtures of smaller-Reynolds-number flows. When combining this ensemble hypothesis with anomalous scaling and universal sets of scaling exponents, we obtain a model that naturally bears various analogies to typical multifractal models. In fact, our model could be considered a variant of the multifractal conjecture. Notably, since small-scale turbulence is represented by fully resolved DNS of Navier-Stokes turbulence, we can make concrete predictions of arbitrary small-scale quantities without further modeling assumptions.

The range of possible applications is large. The power of our method comes from the fact that it is very easy to apply, as it requires only a standard solver for the Navier-Stokes equation and modest computational resources. For any statistics that are accessible in simulations, our method allows to extrapolate to higher Reynolds number in a way consistent with multifractal scaling. We demonstrated this at the example of complex gradient statistics, but in general we expect the predictive power of the model to extend also to short-range spatial and short-time temporal correlations. As an example, a very interesting application would be to add particles to the ensemble simulations in order to study the impact of high-Reynolds-number turbulence on various Lagrangian statistics. As long as the short-time statistics of the flow are the relevant factor for the considered particle statistics, we can hope that our method will give good predictions. Moreover, we here focused on modeling homogeneous, isotropic turbulence. However, if small-scale statistics are universal~\cite{BuariaPRF2025}, then our methodology can be adapted to more realistic, inhomogeneous flows, too. In the simplest form, the $\lambda$ distribution could be varying in space, adapting to the different flow properties. More advanced ensemble models could use ensemble simulations with varying forcing schemes.

On the theoretical side, the insight that the ensemble hypothesis~\eqref{eq:ensemble_hypothesis} immediately leads to results like the Nelkin relation~\eqref{eq:nelkin_rhom} from the multifractal literature is very promising. Exploring the theoretical implications of the ensemble hypothesis for more complex statistical quantities will be the subject of future work. Since~\eqref{eq:ensemble_hypothesis} describes all small-scale statistics in a unified fashion, it would also be very interesting to establish contact with the Navier-Stokes equations for more theoretical insights.

In recent years, there have been discussions about the limitations of the multifractal approach and how to overcome them~\cite{ChenPRL1997a,DhruvaPRE1997,GrossmannPoF1997,ShenPoF2002,GotohPoF2002,GrauerNJP2012,ElsingaJFM2017,IyerPRF2020,ElsingaPRSMPES2020,BuariaPRF2020,BuariaPRL2022,BuariaPRL2023,ElsingaJFM2023}. Since our model is closely related to multifractals, these discussions can give immediate inspiration on how to extend our approach. First, it has been argued that the anomalous scaling exponents are not constant within the Reynolds number ranges currently available to simulations and that they only become constant at much higher Reynolds number~\cite{ElsingaJFM2017,ElsingaPRSMPES2020,BuariaPRF2020,BuariaPRL2022,ElsingaJFM2023}. Adapting our method for varying scaling exponents could further improve the predictive accuracy of the model. Second, there is ample evidence that the scaling exponents of transverse velocity increments at high order differ from the longitudinal ones~\cite{ChenPRL1997a,DhruvaPRE1997,GrossmannPoF1997,ShenPoF2002,GotohPoF2002,GrauerNJP2012,IyerPRF2020,BuariaPRL2023}. Our method in its current form predicts these exponents to be equal instead (due to the same physical dimensions). This is likely the result of assuming a single scalar cascade process (of energy), which is typical for most multifractal models. However, there are many possible extensions to our ensemble approach that may resolve this issue. While we here only varied the energy injection rate across the ensemble, it is conceivable to add additional variations to the forcing. For example, this could be temporal variations, introducing anisotropy, or controlling for helicity injection. Any such changes to the forcing introduce another degree of freedom to the ensemble, which may then allow to incorporate joint cascades with multiple spectra and thus independent sets of scaling exponents~\cite{MeneveauPRA1990,BuariaPRL2023}. In this sense, our ensemble paradigm is not only a way to make concrete predictions of small-scale statistics with minimal assumptions, but it also offers a very practical way to explore the path beyond traditional multifractal modeling.

\section*{Acknowledgments}
We greatly appreciate Dhawal Buaria for providing reference DNS data from his recent publications and for giving helpful advice on the manuscript. We thank Cristian C.\ Lalescu for his technical support. We thank Gabriel Brito Apolin\'ario for providing detailed feedback on the manuscript and we thank Gabriel Brito Apolin\'ario, Tim Niemeyer and Arthur G.\ Peeters for interesting and helpful discussions in the course of this project. GitHub Copilot was used during code development.

\begin{figure*}[ht]
  \includegraphics{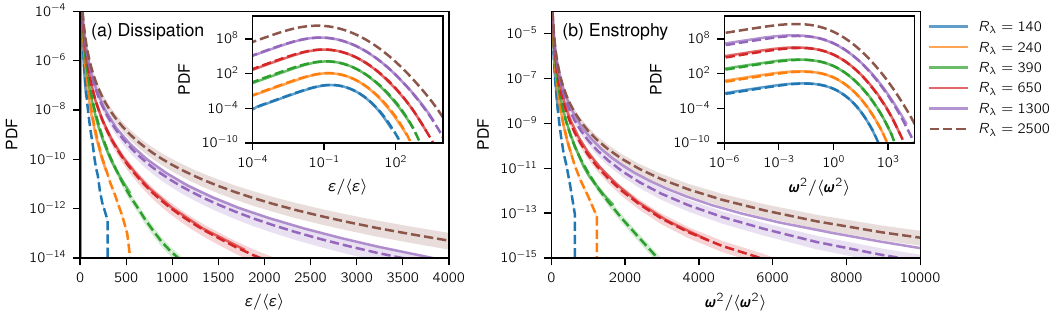}%
  \caption{PDFs of extreme dissipation (a) and enstrophy (b). As in figure~\ref{fig:extreme_gradients}, we compare DNS data from refs.~\cite{BuariaNJP2019,BuariaPRL2022} (solid) to our method (dashed) using the \texttt{ens256} ensemble with refined spatial resolution. However, we replace the Gaussian distributions for $\lambda$~\eqref{eq:gaussian_lambda_pdf} by the truncated ones~\eqref{eq:truncated_gaussian} while keeping $c=0.023$. The shaded areas correspond to predictions based on values of the parameter $0.021 \leq c \leq 0.025$. The insets show the full range of the PDFs in a log-log representation (vertically shifted for clarity).}
  \label{fig:extreme_gradients_truncated}
\end{figure*}
\begin{figure}[ht]
  \includegraphics{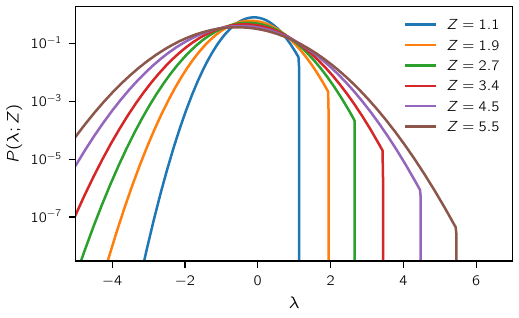}%
  \caption{Truncated Gaussian distributions for $\lambda$ according to~\eqref{eq:truncated_gaussian}, for $c=0.023$ and with $Z$ values corresponding to the ones used in figure~\ref{fig:extreme_gradients_truncated}.}
  \label{fig:lambda_pdfs_truncated}
\end{figure}
\section*{Funding}
This project has received funding from the European Research Council (ERC) under the
European Union's Horizon 2020
research and innovation programme (Grant agreement No.\ 101001081).

The authors gratefully acknowledge the scientific support and HPC resources provided by the Erlangen National High Performance Computing Center (NHR@FAU) of the Friedrich-Alexander-Universit\"at Erlangen-N\"urnberg (FAU) under the NHR project \texttt{b159cb}. NHR funding is provided by federal and Bavarian state authorities. NHR@FAU hardware is partially funded by the German Research Foundation (DFG) -- 440719683.

\appendix

\section*{Appendix: Truncated-Gaussian distribution for $\lambda$} \label{sec:truncated_gauss}
Throughout this paper, we focused on the log-normal model, i.e., the quadratic parameterization~\eqref{eq:quadratic_scaling_exponents} of the $\zeta_p$ exponents. However, it is known that for a bounded velocity field, the $\zeta_p$ exponents must be monotonous~\cite{Frisch1995}, inconsistent with the quadratic form. In fact, recent DNS results~\cite{IyerPRF2020,BuariaPRL2023} suggest that the $\zeta_p$ exponents saturate at large values of $p$, i.e.,
\begin{align}
\zeta_p \xrightarrow[]{p\to\infty} \mathrm{const}\,.
\end{align}
Taking a $p$-derivative of~\eqref{eq:exponents_KZ_1} shows that this corresponds to
\begin{align}
  \dod{}{t} K(t; Z) \xrightarrow[]{t\to\infty} Z \,.
\end{align}
For the $\lambda$ distribution, this saturation of $\tod{}{t}K$ at large $t$ can be achieved by a sharp cutoff of its support. Specifically, if the support of the $\lambda$ distribution reaches up to $\lambda_\mathrm{max}(Z)$, then we have
\begin{align}
  \dod{}{t} K(t; Z) &= \frac{\int\dif\lambda\,P(\lambda; Z) \lambda e^{\lambda t}}{\int\dif\lambda\,P(\lambda; Z) e^{\lambda t}}
  \xrightarrow[]{t\to\infty} \lambda_\mathrm{max}(Z)\,.
\end{align}
So saturation of the exponents corresponds to $\lambda_\mathrm{max}(Z) = Z$ for all $Z$. Indeed, we find this behavior numerically for the $\lambda$ distribution inferred from the saturating $\zeta_p$ parameterization by Sreenivasan \& Yakhot~\cite{sreenivasan2021} (used for figure~\ref{fig:zetap_choices}).

For the Gaussian distribution~\eqref{eq:gaussian_lambda_pdf} used throughout this paper, we can achieve this behavior by a truncation:
\begin{align} \label{eq:truncated_gaussian}
  P(\lambda; Z) &= N(Z) \exp\left(-\frac{(\lambda - \mu(Z))^2}{18cZ}\right) \Theta(Z - \lambda)\,.
\end{align}
Here, $\Theta(x)$ is the Heaviside function, and $N(Z)$ and $\mu(Z)$ are chosen such that the PDF is normalized and satisfies energy conservation~\eqref{eq:mean_injection_constraint}. We keep $c=0.023$ as above. Figure~\ref{fig:extreme_gradients_truncated} shows the PDFs of extreme dissipation and enstrophy as in figure~\ref{fig:extreme_gradients} but using the truncated Gaussian distribution of $\lambda$ for the model predictions. The agreement is noticeably better, in particular at the smaller Reynolds numbers where the truncation has a stronger effect. The corresponding $\lambda$ distributions are shown in figure~\ref{fig:lambda_pdfs_truncated}. However, note that the distribution~\eqref{eq:truncated_gaussian} does not satisfy the convolution property~\eqref{eq:convolution_property} and thus does not map uniquely to a set of scaling exponents through~\eqref{eq:exponents_KZ_1}. Nevertheless, the example shows that even better results can be achieved by adjusting the $\lambda$ distribution.

\bibliography{../../refs}

\end{document}